\documentstyle[prb,twocolumn,epsf,amssymb,floats,aps]{revtex}
\begin{document}
\draft

\twocolumn[\hsize\textwidth\columnwidth\hsize\csname @twocolumnfalse\endcsname

\title{N{\'e}el to spin-Peierls transition in the ground state of a quasi-1D Heisenberg model coupled to bond phonons}

\author{Pinaki Sengupta} 
\address{Department of Physics, University of California, Davis, CA 95616}

\date{\today}

\maketitle

\begin{abstract}
The spin-Peierls transition in the ground state of a quasi-one-dimensional
spin-$1\over 2$ Heisenberg model coupled to adiabatic bond phonons is studied
using a quantum Monte Carlo (QMC) method. The transition from a gapless 
N{\'eel} state to a spin-gapped Peierls state is explored in the parameter
space spanned by the spatial anisotropy and the strength of spin-lattice 
coupling. It is found that for any finite inter-chain coupling, the 
transition to a dimerized Peierls ground state occurs only when the 
spin-lattice coupling exceeds a finite, non-zero critical value. This is
in contrast to the pure 1D model, where adiabatic phonons lead to a 
dimerized ground state for any non-zero spin-phonon coupling. The
phase diagram in the above parameter space is mapped out. No evidence is
found for a region of co-existing long range magnetic order and dimerization.
The vanishing of the N{\'e}el order occurs simultaneously with the 
setting in of the dimerization.
\end{abstract}

\vskip2mm]

\section{Introduction}
When coupled to an elastic lattice, a spin-{$1\over 2$} Heisenberg chain
is unstable towards a dimerization of the lattice. The cost in elastic 
energy due to a distortion $\delta$ of the lattice
($\sim \delta^2$) is compensated for by the gain in the magnetic energy 
($\sim \delta^{4/3}$), and the ground state is stabilized for a lattice
with non-zero dimerization\cite{pytte,cross} and a finite spin gap. 
Such a transition is known as the spin-Peierls (SP) transition -- in 
analogy with the conventional
Peierls transition in a one-dimensional (1D) metal. The SP transition has been 
extensively studied and is believed to be fairly well understood in 1D.
\cite{pytte,cross,uhrig1,gros,wellein,aws-sp,bursill,weisse,raas} 
In the adiabatic limit, any arbitrarily small coupling to an elastic lattice 
leads to a dimerized ground state. On the other hand, 
quantum lattice fluctuations destroy the dimerization for small spin-phonon
couplings and/or large bare phonon frequencies.\cite{aws-sp,bursill} 
The transition to a dimerized 
(Peierls) state occurs only when the spin-phonon coupling exceeds a finite,
non-zero critical value (that depends on the bare phonon frequency). This is 
true for both the spin-${1\over 2}$ Heisenberg and $XY$ models with 
spin-phonon coupling. The discovery\cite{hase} of the quasi-1D inorganic 
spin-Peierls compound CuGeO$_3$ has led to
a resurgence in the study of spin-Peierls(SP) transition in low-dimensional
spin models. The properties of CuGeO$_3$ have been widely studied
within the framework of a 1D spin-${1\over 2}$ Heisenberg model coupled to
phonons, with an additional frustrated next-nearest-neighbor interaction.\cite{riera}

Fewer studies exist for the spin-Peierls transition in two
dimensions (2D) and the mechanism of the transition is not as well established
as in 1D. 
Although the study of the Peierls transition in 2D or quasi-1D systems 
presents a technically much
more difficult challenge, it also allows for a more realistic modeling of
real materials -- for CuGeO$_3$, the strength of inter-chain 
coupling is estimated to be $J_\perp/J\approx 0.1$ 
experimentally\cite{nishi}, and its importance in any realistic modeling
of CuGeO$_3$ has been widely stressed.\cite{uhrig2,bouzerar}
From a purely theoretical standpoint, the study of spin-Peierls transition
is interesting on its own. Unlike in 1D, the
ground state of the 2D (Heisenberg) system, in the absence of spin-phonon 
coupling, has long range antiferromagnetic (N{\'e}el) order. It is generally
believed that even for adiabatic phonons, the spin-phonon coupling has to
exceed some non-zero critical value for the ground state to develop a 
dimerized pattern with a spin gap. What is the nature of the transition?
Is there a region in the phase space where the ground state has co-existing
dimerization and long range antiferromagnetic order? Furthermore, the 
existence of several different possible dimerization pattern in 2D means that,
in principle, different dimerization patterns can be stabilized for different
values of the parameters. Finally, unlike in 1D, the Peierls phase in 2D 
extends to finite temperatures.

The spatially isotropic 2D spin-$1\over 2$ Heisenberg model with static
dimerization patterns have been studied by several authors.
\cite{if1,if2,katoh,koga,al-omari,sirker} By comparing the ground state 
energies for different patterns, these authors have tried to 
predict the energetically most favored dimerization pattern. Using the same 
strategy, the 2D
tight binding model with bond-distortions\cite{ono} and the 2D Peierls-Hubbard 
model\cite{tang,sumit,zhang} have also been studied. Since in the limit of large
on-site repulsion, $U$, the Hubbard model at half-filling reduces to the
Heisenberg model, results from the Peierls-Hubbard model (in the limit of large 
$U$) are expected to be applicable to the present discussion.
Unfortunately, there is no consensus among the different studies as to the nature
of the dimerization pattern in the ground state. For the Peierls-Hubbard model 
at half-filling, Tang and
Hirsch\cite{tang} find a plaquette-like distortion to be energetically
favored in the limit of large $U$. On the other hand, Mazumdar\cite{sumit} have 
argued that the minimum energy ground state has a ``stairlike'' dimerization 
pattern -- corresponding to a wave vector $\bf{Q}=(\pi,\pi)$. Zhang and
Prelov\v{s}ek\cite{zhang} agree with a dimerization pattern with $\bf{Q}=(\pi,\pi)$,
but conclude that the ground state has dimerization only along one of the
axes (staggered dimerized chains -- similar to the pattern considered here).
For the Heisenberg model with static dimerization, Al-Omari\cite{al-omari}
has concluded that the ground state energy is minimized for a plaquette-like 
distortion of the lattice which agrees with the conclusion of Tang and Hirsch. 
On the other hand, Sirker {\it et.al.}\cite{sirker} find that linear spin wave 
theory (LSWT) predicts a stairlike dimerization pattern to be most favored, 
in agreement with Mazumdar. Using LSWT, Sirker {\it et.al.} also find finite 
regions in the parameter space with co-existing long range magnetic order and 
non-zero dimerization. However, as pointed out by the authors, results obtained 
from LSWT are not reliable at large values of dimerization. The effects of 
inter-chain coupling was considered early on
by Inagaki and Fukuyama\cite{if1,if2} who studied a quasi-1D system 
of coupled spin-${1\over 2}$ Heisenberg chains with a fixed dimerization 
pattern corresponding to a wave vector ${\bf Q}=(\pi,0)$. By treating the 
inter-chain coupling in a mean-field theory, they were able map out the 
ground state phase diagram\cite{if1} and study the finite temperature 
transition\cite{if2}. Later, Katoh and Imada\cite{katoh} studied in
detail the nature of the transition. More recently, the effects of impurities 
have been studied in this model (once again with a mean-field treatment 
of the inter-chain coupling) which revealed a region of co-existing Peierls 
and antiferromagnetic orders.\cite{khomskii,saito,affleck,dobry,melin1,melin2} 
In addition to this, the quasi-1D $XY$ model with a ${\bf Q}=(\pi,\pi)$
dimerization pattern has recently been studied\cite{ji,yuan} by an extension 
of the Jordan-Wigner transformation in 2D.\cite{wang1,wang2,azzouz} The effects
of quantum phonons in the isotropic 2D model have also been studied,\cite{low}
where the authors find that there is
no evidence of a transition to the Peierls state for a wide range of values
of the bare phonon frequency and the spin-phonon coupling. This is 
consistent with previous finding\cite{rts} for the same model 
that the spin wave spectrum along the Brillouin zone boundary is qualitatively
similar to that for the pure Heisenberg model.

The present work aims to investigate the SP transition in a spin-${1\over 2}$
quasi-1D Heisenberg antiferromagnet coupled to static $\bf{Q}=(\pi,\pi)$ bond 
phonons with varying inter-chain coupling. The elastic energy due to the 
lattice distortion is considered as well as the change in magnetic energy.
The nature of the ground state is explored for different values of the
inter-chain coupling and the elastic constant associated with the bond 
distortions and the ground state phase diagram in this phase space is
mapped out.

The rest of the paper is organized as follows. In Section II, the model 
Hamiltonian and the Stochastic Series Expansion (SSE) QMC method 
used to study it are introduced. The results of the simulations are presented 
in Section III. Section IV concludes with a summary of the results.

\begin{figure}
\centering
\epsfxsize=8.3cm
\leavevmode
\epsffile{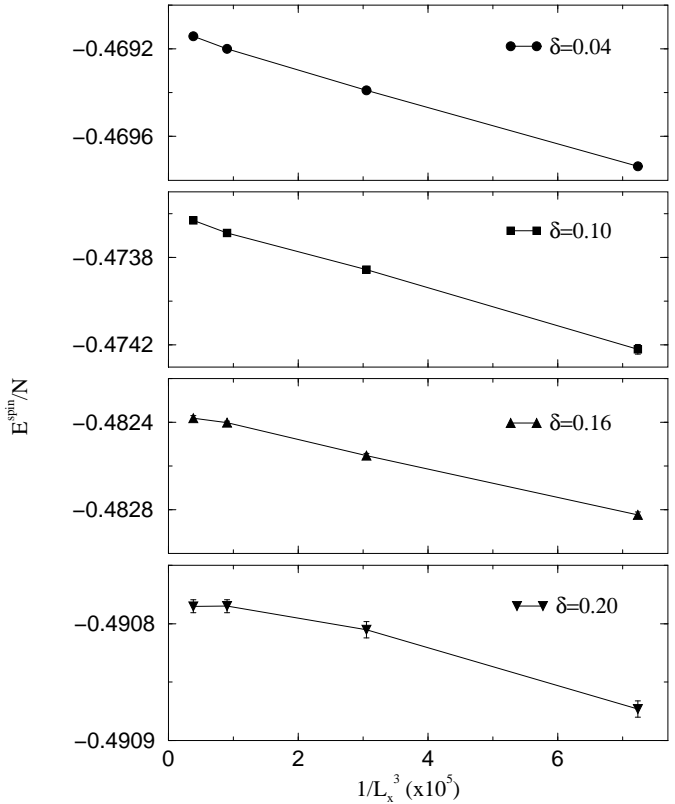}
\vskip1mm
\caption{The spin part of the ground state energy as a function of
the system size for four values of $\delta$, and $\alpha$=0.25.} 
\label{fig:econv}
\end{figure}

\section{Model and Simulation techniques}

A quantum Monte Carlo (QMC) method has been used to study a quasi-1D Heisenberg 
model with spin-phonon coupling. The model is given by the Hamiltonian
\begin{eqnarray}
H &=& J\sum_{i,j}(1+\lambda u_{i,j}){\bf S}_{i,j}\cdot{\bf S}_{i+1,j} + {1\over 2}K\sum_{i,j}u_{i,j}^2 \nonumber \\
  & & +J_\perp\sum_{i,j}{\bf S}_{i,j}\cdot{\bf S}_{i,j+1},
\end{eqnarray}
where $J_\perp$ is the inter-chain coupling, $\lambda$ is the strength of
the spin-phonon coupling (restricted to be only along the chains), $u_{i,j}$'s
are the distortions of the bond lengths and $K$ is the elastic constant for 
the distortions. Following the experimentally observed\cite{tranquada} dimerization pattern
in CuGeO$_3$, the bond length distortions are chosen to be of the form
\[
u_{i,j}=(-1)^{i+j}\delta
\]
This amounts to choosing a fixed dimerization pattern corresponding to the
wave-vector ${\bf Q}=(\pi,\pi)$. As in previous works, the bond distortions 
can be rescaled by the spin-phonon coupling strength, $\lambda$. This
reduces the Hamiltonian to
\begin{eqnarray}
H &=& \sum_{i,j}(1+(-1)^{i+j}\delta ){\bf S}_{i,j}\cdot{\bf S}_{i+1,j} + N\delta^2/2\zeta \nonumber \\
  & & +\alpha\sum_{i,j}{\bf S}_{i,j}\cdot{\bf S}_{i,j+1},
\label{eqn:H}
\end{eqnarray}
where $\zeta={\frac{\lambda^2J}{K}}$, $\alpha=J_\perp/J$ and $N$ 
is the size of the lattice. The static approximation for the 
displacements makes the computational task easier. The following
strategy is adopted. The simulations are carried out for the spin 
variables for  
several different $\{\alpha,\delta\}$ parameter sets. This produces
the spin energy of the system as a function of $\delta$ for a fixed
$\alpha$. Next the elastic energy with a particular $\zeta$ is added and 
the total energy is minimized to obtain the value of the ground state 
distortion
for the given set of parameters $\{\alpha,\zeta\}$. This is repeated for 
different sets of $\{\alpha,\zeta\}$ to obtain the ground state phase
diagram in the parameter space spanned by $\alpha$ and $\zeta$.

The stochastic series expansion (SSE) QMC method has been used to study
the present model. The SSE method \cite{aws1,aws2,sseloop}
is a finite-temperature quantum Monte Carlo method based on importance
sampling of the diagonal elements of the Taylor expansion of $e^{-\beta H}$, 
where $\beta$ is the inverse temperature $\beta=J/T$. Ground state expectation
values can be obtained using sufficiently large values of $\beta$, and there
are no approximations beyond statistical errors. Using the recently
developed ``operator loop update'',\cite{sseloop} it is possible to 
explore the phase space of the Hamiltonian (\ref{eqn:H}) in an efficient 
manner. 

\begin{figure}
\centering
\epsfxsize=8.3cm
\leavevmode
\epsffile{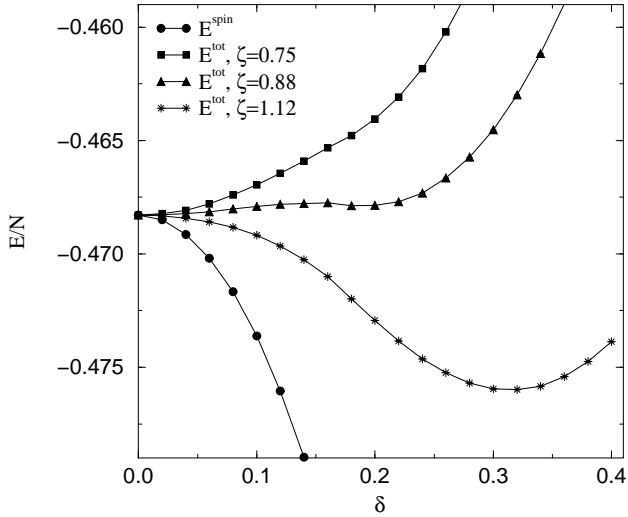}
\vskip1mm
\caption{The spin part and the total ground state energy as a function of
the bond distortion for 3 different values of the elastic constant, $\zeta$,
and fixed $\alpha$=0.25. The system size is $N$=64x16.} 
\label{fig:etot}
\end{figure}

\section{Results}

For a system of weakly coupled Heisenberg chains, it was shown recently
that the estimates for various observables for a spatially anisotropic
system depend non-monotonically on the system size for square ($L_x = L_y$)
geometry.\cite{multichain} One has to go to rectangular ($L_x \neq L_y$)
geometries to obtain 
monotonic behavior of the numerical results for extrapolating to
the thermodynamic limit. Similar effects are expected in
the present model for $\alpha \ll 1$. Hence rectangular lattices 
with the aspect ratio $L_x = 4L_y$have been studied, with $L_x=16-64$. An inverse 
temperature of $\beta=8L_x$ was found to be sufficient for the observables
to have converged to their ground state values. Nine different
values of the inter-chain coupling were considered, $\alpha=0.01,\ldots, 1.00$.

\begin{figure}
\centering
\epsfxsize=8.3cm
\leavevmode
\epsffile{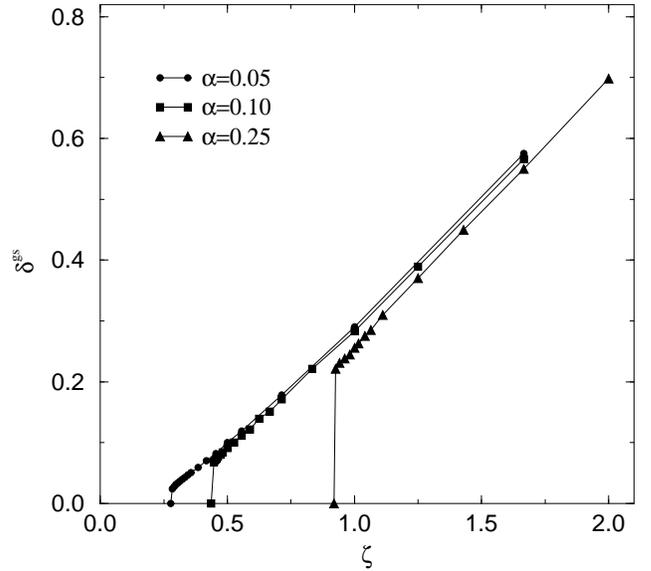}
\vskip1mm
\caption{The equilibrium ground state bond distortion as a function of the
spin-lattice coupling for different values of the inter-chain coupling.} 
\label{fig:deq}
\end{figure}

Fig.\ref{fig:econv} shows the finite-size dependence of the QMC data 
for the spin part of the ground state energy for four different values 
of the bond length distortion, $\delta$. The values of $\delta$ chosen 
correspond to a N{\'e}el ground state ($\delta$=0.04, 0.10), a Peierls 
ground state ($\delta$=0.20), and a ground state close to the critical 
$\delta$ for N{\'e}el to spin-Peierls transition in the pure spin system 
(the phase transition in the spin part of the Hamiltonian (\ref{eqn:H})
is discussed at the end of the current section). 
The leading order finite-size correction to the ground state energy is
seen to be $\sim 1/L_x^3$ for small values of $\delta$ corresponding to
a N{\'e}el-ordered ground state. This is similar to that observed for the 
pure 2D Heisenberg model.\cite{aws2} On the other hand, for a spin-gapped
ground state ($\delta$=0.20), the energy scales exponentially
with system-size. Close to the critical point, extrapolation to the 
thermodynamic limit becomes difficult due to cross-over effects. 
Instead, the data from the largest system size studied have been used 
to map out the phase diagram. Fortunately, the data for the largest 
system sizes studied are found to be well 
converged -- the fractional difference in the energy for the two largest
system sizes studied is $\sim 10^{-5}$. This observed convergence allows 
for reliable estimation of ground state properties in the thermodynamic 
limit based on the data from the largest system size -- any finite-size effects 
on such estimates are expected to be small.

The strategy implemented to extract the ground state bond distortion is
qualitatively demonstrated in fig.\ref{fig:etot}. The total ground state
energy is obtained by adding the 
elastic energy contribution to the spin part of the energy obtained from 
the simulations. The plot shows the spin part of the energy, 
as well as the total ground state energy as a function of the bond length
distortion, $\delta$, for three different values of the elastic
energy constant, $\zeta$, and a fixed value of the inter-chain coupling
($\alpha$=0.25). For large $\zeta$, the total energy is 
minimized for a non-zero value of the bond length distortion, $\delta$, 
whereas for small $\zeta$, a uniform ground state with $\delta$=0 is 
energetically favored. The behavior of the total energy near the critical 
$\zeta$ is also shown. The ground state distortion in bond length is 
obtained by numerically differentiating the total energy data and 
solving for $\frac{\partial {\mbox E^{tot}}}{\partial\delta}|_{\delta^{gs}}=0$.
In principle, one can also fit a polynomial to the QMC data to get 
${\mbox E^{spin}}(\delta)$ and add to it the elastic energy term to get
${\mbox E^{tot}}(\delta)$. Then the ground state distortion can be obtained
as a continuous function of $\delta$ by solving 
$\frac{\partial {\mbox E^{tot}}}{\partial\delta}|_{\delta^{gs}}=0$.
However, in practice, the earlier method is found to be more reliable
because of the uncertainty in the order of the polynomial fit.

Fig.\ref{fig:deq} shows the equilibrium distortion in the bond lengths in
the ground state of the system as a function of the elastic energy parameter,
$\zeta$, at fixed values of $\alpha$, obtained as described above. For small 
values of $\zeta$, the tendency
towards dimerization is suppressed, and a uniform (N{\'e}el ordered) ground 
state with uniform bond lengths is stabilized. As $\zeta$ is increased above 
a critical value, $\zeta_c$, there is a transition to a ground state with a finite,
non-zero dimerization. For $\zeta > \zeta_c$, the equilibrium ground state 
distortion increases monotonically with $\zeta$. The critical value $\zeta_c$
required for a transition to a dimerized ground state depends on the value
of the inter-chain coupling, $\alpha$ and increases with increasing $\alpha$.
The data indicate that the transition is a discontinuous (first order) 
quantum phase transition for any finite value of inter-chain coupling.

\begin{figure}
\centering
\epsfxsize=8.3cm
\leavevmode
\epsffile{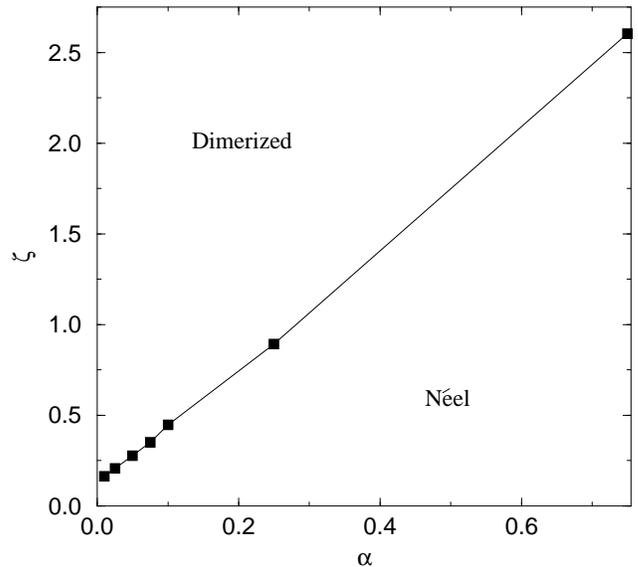}
\vskip1mm
\caption{The ground-state phase diagram in the parameter space of the
inter-chain coupling and the spin-lattice coupling strength.} 
\label{fig:phase}
\end{figure}

The results from fig.\ref{fig:deq} are summarized in fig.\ref{fig:phase}, which
shows the phase diagram for the system in the phase space spanned by the 
parameters $\zeta$ and $\alpha$. For small $\zeta$ and/or large $\alpha$, the
ground state of the system is N{\'e}el ordered with zero bond distortion,
while for large $\zeta$ and/or small $\alpha$, the ground state is 
dimerized with a finite spin gap. The phase boundary is roughly linear over the
range of values of $\alpha$ studied. However, it is well known that in the
1D limit ($\alpha = 0$), the critical elastic constant should vanish ($\zeta_c=0$).
Unfortunately, the numerical data becomes too noisy in this limit for an accurate
estimate of the critical coupling. For the quasi-1D $XY$ model, it was shown that 
$\zeta_c$ goes as $-1/{\mbox{ln}}\alpha$ for $\alpha \rightarrow 0$. On the other hand, 
results for a system of coupled (undimerized) Heisenberg chains show that the
magnetization vanishes as a power law in $\alpha$ with a logarithmic correction.\cite{multichain} 
A similar behavior is expected for the critical coupling in the present model.

\begin{figure}
\centering
\epsfxsize=8.3cm
\leavevmode
\epsffile{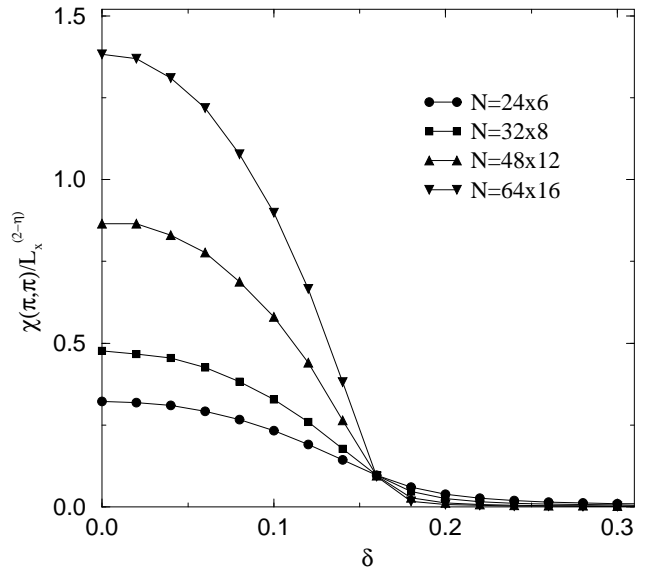}
\vskip1mm
\caption{The ground-state static staggered magnetic susceptibility 
as a function of bond distortion for a fixed value of the inter-chain
coupling, $\alpha=0.25$.}
\label{fig:xs}
\end{figure}

This brings up the interesting possibility of having a region in the 
($\zeta,\alpha$) parameter space where the ground state has co-existing
N{\'e}el order and non-zero dimerization. Such co-existence has been shown
to exist in the presence of doping.\cite{khomskii,saito,affleck,dobry,melin1,melin2} 
For this purpose, the static spin susceptibility defined as
\begin{equation}
\chi({\bf q})={1\over N}\sum_{\langle i,j \rangle}e^{i{\bf q}\cdot ({\bf r}_j-{\bf r}_j)}\int^{\beta}_0d\tau\langle S^z_j(\tau)S^z_i(0) \rangle.
\end{equation}
have been studied for the spin part of the Hamiltonian (\ref{eqn:H})(without the elastic 
energy term).
\begin{eqnarray}
H &=& \sum_{i,j}(1+(-1)^{i+j}\delta ){\bf S}_{i,j}\cdot{\bf S}_{i+1,j} + \nonumber \\
  & & +\alpha\sum_{i,j}{\bf S}_{i,j}\cdot{\bf S}_{i,j+1},
\label{eqn:Hspin}
\end{eqnarray}
If the ground state has long range antiferromagnetic order, the staggered 
(${\bf Q}=(\pi,\pi)$) susceptibility scaled by the system size
($\chi(\pi,\pi)/N$),
for a finite system will increase with increasing system size, diverging
in the thermodynamic limit. On the other hand, if the ground state has
a finite spin gap, $\chi(\pi,\pi)/N$ will vanish in the limit
of infinite system size. This qualitative criterion can be expressed in a
more quantitative manner by noting that the ground state of the above 
Hamiltonian undergoes 
a continuous transition from a N{\'e}el ordered state with long range 
antiferromagnetic order to a spin-gapped, dimerized phase with no long range 
magnetic order as $\delta$ is increased beyond a finite, non-zero critical 
value, $\delta^*$, that depends on the inter-chain coupling $\alpha$. The 
transition belongs to the universality class of the 3D Heisenberg model\cite{chn}.
Finite-size scaling\cite{barber} predicts that for such a transition, 
the finite-size
susceptibility at the critical $\delta$ scales with the system size as
\[
\chi(L_x)\sim L_x^{2-\eta},
\]
for a rectangular lattice of dimension $N=L_x$x$L_y$.
This implies that on a plot of $\chi(\pi,\pi)/L_x^{2-\eta}$, the curves
for different system sizes will intersect at the critical $\delta$. The
value of the critical exponent $\eta$ is known to a high degree of accuracy
($\eta \approx 0.037$).\cite{campostrini}
Fig. \ref{fig:xs} shows  $\chi(\pi,\pi)/L_x^{2-\eta}$ as a 
function of $\delta$ for a fixed value of $\alpha = 0.25$ for several different
system sizes. For small $\delta$, the scaled susceptibility increases with 
increasing system size, indicating the presence of long range magnetic
order. For larger values of $\delta$, the scaled susceptibility goes to zero
with increasing system size, signaling the opening up of a spin gap. From the 
data, the critical value of $\delta$ is estimated to be 
$\delta^* \approx 0.16$. This value of the bond
length distortion is less than the jump in $\delta$ at the transition point,
$\zeta_c$. This means that for $\zeta > \zeta_c$, the ground state is dimerized
with $\delta^{gs} > \delta^*$ and has no long range magnetic order. On the
other hand, for $\zeta < \zeta_c$, the a uniform ground state is energetically
favored that has zero dimerization ($\delta^{gs}=0$), and long range magnetic
order (the N{\'e}el state). For no values of $\zeta$ is a ground state with
$0 < \delta^{gs} < \delta^*$ ever favored energetically. Hence the transition 
to the dimerized phase is accompanied by the disappearance 
of magnetic order and there is no region of co-existing dimerization and magnetic 
order. This is true for the present model. Whether it is possible to have ground
states with co-existing magnetic order and dimerization in other models 
(with different dimerization patterns) remains to be seen.

\section{Summary}
A quantum Monte Carlo method has been used to study a quasi-1D Heisenberg
model coupled to static bond phonons. Motivated by experimental observations
in the inorganic quasi-1D spin-Peierls compound CuGeO$_3$, the bond 
distortions are restricted to be only along the chains, and a single
dimerization pattern, corresponding to the wave vector 
${\bf Q}=(\pi,\pi)$, is considered. It is found that unlike the pure 1D case,
for any non-zero inter-chain coupling $\alpha$, the transition to a dimerized
spin-Peierls ground state occurs only when the spin-lattice coupling strength,
$\zeta$, exceeds a finite, non-zero critical value, $\zeta_c$, even for the
static phonons considered here. For $\zeta<\zeta_c$, the ground state has
long range N{\'e}el order and zero spin gap, whereas for $\zeta>\zeta_c$,
the ground state develops a finite dimerization accompanied by the
opening up of a spin gap. The transition is found to be a discontinuous
(first order) quantum phase transition. The value of the critical coupling
depends on the strength of the inter-chain coupling, increasing monotonically
with $\alpha$. The phase diagram in the parameter space of $\alpha$ and $\zeta$
is mapped out. Finally it is found that in the present model, the transition 
to the dimerized Peierls state is accompanied by the vanishing of magnetic order, 
and that there is no region of co-existing magnetic order and non-zero 
dimerization. 

\section{Acknowledgments}

The author would like to thank Anders Sandvik and Rajiv Singh for useful
discussions. This work was supported in part by NSF grant number DMR-9986948. 
Simulations were carried out at the IBM SP facility at NERSC, Berkeley. 

\null

\end{document}